\documentstyle[12pt,psfig]{article}
    \date{}
\begin{document}
    
    \title{On Isgur's "Critique of a Pion Exchange Model
           for Interquark Forces"}
    
    \author{L. Ya. Glozman}
    \maketitle
    
    \centerline{\it  Institute for Theoretical Physics, 
    University of Graz,}
    \centerline{\it Universit\"atsplatz 5, A-8010 Graz, Austria}
    
    \setcounter{page} {0}
    \vspace{1cm}
    
    \centerline{\bf Abstract}
    \vspace{0.5cm}
    The conceptual issues of  low-energy baryon physics
    are discussed. In particular, a comparison between the
    naive one gluon exchange model for the interaction between
    constituent quarks in hadrons and the Goldstone boson
    exchange picture is made. The "defects" of the Goldstone
    boson exchange model for  baryons, indicated by Isgur
    \cite{Isgur} are examined in detail. {\it All} of the purported
    ``defects'' are shown to lack a valid basis.

    \bigskip
    \bigskip

    \bigskip
    \bigskip
    
    \noindent
    PACS number(s): 12.39.Jh, 14.20.-c
    \small

    \newpage
    \section{Introduction}

The recent ``critique'' \cite{Isgur} of the Goldstone
boson exchange (GBE) model \cite{GR,GPVW} for the baryon
spectra contains a number of strong and at the same time
unsubstantiated statements.
Given
the author's "silence is consent" \footnote {
See the first version of the
paper which has become widely known and which can be
extracted from the LANL e-print server as version 1.} and an
increasing pressure from the community, 
a rebuttal has become unavoidable. In this rebuttal the 
structure of  Isgur's paper \cite{Isgur} will be
adhered to and 
his "catalogue of criticisms" will be examined. An updated
version of the GBE model for the baryon spectrum is 
available in ref.\cite{PANIC}.
We discuss conceptual issues related to a question
of paramount importance: which physics, inherent in QCD,
is responsible for the nucleon (baryon) mass and its 
low-energy properties and how this physics is connected
with the observed baryon spectra.
\\

In the introduction to the first variant of his paper
Isgur \cite{Isgur} questions the superiority of
the GBE model for solving the problem of the spectral
ordering in light and strange baryons, and argues that the Coulomb
component of the one gluon exchange (OGE) interaction naturally
leads the positive parity state $N(1440)$
to be the lowest one among
positive parity  $N=2$ band. This issue is dropped from the final
variant, but as the problem of the relative ordering of the
lowest positive-negative parity states is the key question 
for deciding which physical picture is responsible for baryon 
(nucleon) masses,
we will shortly address it here.
\\

In a model with a monotonic effective confining interaction
between quarks in light and strange baryons, which is flavor-
and spin-independent, and assuming that there are no residual
interactions, the spectrum of the lowest lying baryons should be
arranged into successive bands of positive and negative parity
(Fig. 1). Empirically, however, the lowest excited levels in the spectra
of nucleon, the $\Delta$ - resonance and $\Lambda$-hyperon, which are
shown in Fig. 2, look quite different. It follows
that a picture, in which all other possible interactions are treated as
only residual and weak  and represent only a perturbation
cannot be correct.\\

\begin{figure}
\psfig{file=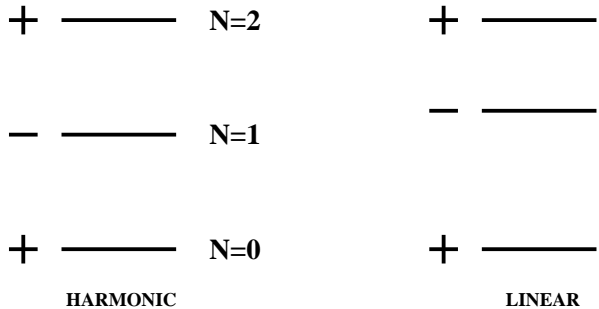}
\caption{A sequence of positive and negative parity levels
with harmonic and linear confining interactions.}
\end{figure}

In the other extreme case, with  a very strong Coulomb
interaction between quarks and without any confining force at all,
the lowest excited positive and negative parity states should be
degenerate in all flavor parts of the spectrum, as in the hydrogen atom.
Experimentally, however, the positive parity state
N(1440) lies $\sim 100$
MeV below the negative parity multiplet N(1535)-N(1520), 
on the one hand, but on the other hand the lowest positive
parity state in the $\Lambda$ spectrum lies 100 - 200 MeV
above the lowest negative parity doublet (Fig. 2).
This rules out the hypothesis of a dominant Coulomb interaction.
In addition, a model with no
confining interaction, that relies exclusively on the Coulomb part
of OGE, fails for the spectra of all other low-lying
baryons. Such a model cannot provide the required 
500 MeV gap between the ground
state baryons and the first negative parity excitation band. As soon
as a confining interaction is added, irrespective of whether
harmonic,
linear or some other monotonic functional form, 
the Roper resonance (and its counterparts in other flavor parts of
the spectrum) falls $\sim 100 - 300$
MeV above the negative parity multiplet, a result which is well
known from many exact 3-body calculations, see e.g. \cite{CKP,SB,KN}, 
including
those of Isgur \cite{CI}.\footnote{In Isgur's papers with Karl \cite{IK},  
the positive and negative parity states
are treated separately in different papers, and a very strong color
coupling constant - larger than 1 - is needed to fit the $N- \Delta$
mass splitting, which is incompatible with the perturbative
treatment of QCD,
and  a huge anharmonic
"correction" is introduced by hand in order to cure the positive parity
states. This is the salient point as it is assumed in the Isgur-Karl
model that the anharmonicity, i.e. difference between the harmonic
interaction and the linear + Coulomb interaction could shift the
Roper strongly down. This is excluded in exact three-body calculations
\cite{CKP,SB} and also at the theorem level - see \cite{KN} (and references cited therein).} 
It then follows that a combined model, relying
on both the confinement potential and color-Coulomb component of
one gluon exchange cannot explain the experimentally observed pattern.\\

\begin{figure}
\psfig{file=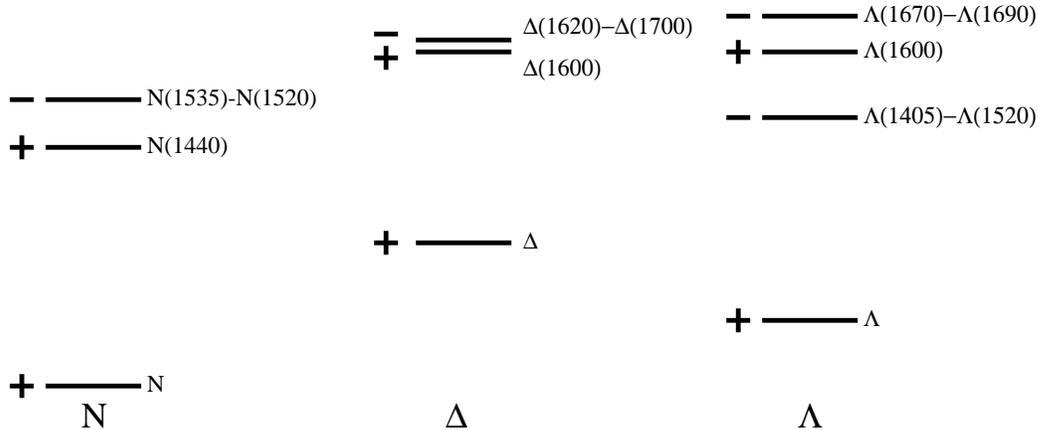}
\caption{Low-lying spectra of nucleon, $\Delta$-resonance and 
$\Lambda$-hyperon.}
\end{figure}

The next important issue is how these spectra change
when perturbed by the color-magnetic component of 
OGE. To leading order and when one ignores the spatial
dependence of the  color-magnetic interaction
and assuming the $SU(3)_F$ limit, its contribution
is determined exclusively by the spin structure of the zero-order
baryon wave function, which is prescribed by the corresponding
Young diagram.
 This spin structure is unambiguously
determined by the total spin of three quarks. This spin is the
same, $S=1/2$, for all baryons in $N$ and $\Lambda$ spectrum,
depicted in Fig. 2. This then implies that such a spin-spin force,
which is not sensitive to the flavor of quarks, cannot modify
the ordering of the states, suggested by the confinement + Coulomb
interaction. If one takes into account a spatial dependence of the
color-magnetic interaction
as well as the $SU(3)_F$ breaking, its contribution to the positive and
negative parity states will be slightly different, because of
the different radial structure of these baryons. Nevertheless, a 
first order perturbation calculation or nonperturbative calculations
\cite{CKP,SB,CI} reveal\footnote{Note that the theoretical predictions
for all positive and all negative parity states are shifted down and up, respectively, in figures of the ref. \cite{CI}, so after a reconstruction
of actual picture as it follows from the Tables of that paper it is
very difficult to conclude that a reasonable description of the spectra
has been obtained!},
 that the departures from the pattern of Fig. 1
are small.  But even more seveer is
the constraint from the $\Delta$ spectrum. In this case the
color-magnetic interaction shifts the $N=2$ state $\Delta(1600)$ 
($S=3/2$) up, but not down, with respect to the negative parity $N=1$
pair $\Delta(1620) - \Delta(1700)$ ($S=1/2$) !
All these facts rule out the perturbative gluon exchange plus confinement
picture as a physical mechanism for the generation of
light baryon mass.  These spectra obviously point to explicit flavor
dependence in the underlying dynamics. The GBE force, which is 
explicitly flavor dependent, very naturally explains this as well as 
several other apparent puzzles\footnote{In fact, one needs only two 
parameters to explain the
pattern of Fig. 2 in the most simple form of the GBE model
\cite{GR}: the strength of the GBE-like interaction
and the strength of the effective confining interaction \cite{PANIC}.
This is in contrast to quite
 a misleading counting of the number of free parameters as
done in a recent review \cite{Capstick}.}\cite{GR,GPVW,PANIC}.\\

But in a sense more important is the conceptual inadequacy of 
the simplistic OGE 
model. This model invokes constituent quarks as particles
with constant mass, without any
attempt to understand an essense of these objects, which is very
different from that of the light current quarks of QCD. 
The constituent quark can be introduced as a quasiparticle
in the Bogoliubov or Landau sense stemming from the dynamical
chiral symmetry breaking in QCD and related to the quark
condensate in the QCD vacuum (in fact one cannot obtain a nonzero
condensate without  a dynamical mass of quarks).
Such a dynamical mass is indeed observed on the lattice at
small momenta where the nonperturbative phenomena become
crucial \cite{Aoki,Sku}.
 This physics is well known \cite{NJL}
and it is a common theme in all strongly interacting many fermion systems,
to which QCD belong. The chiral symmetry breaking and the dynamical
mass generation are inherently nonperturbative phenomena and
cannot be addressed within perturbation theory. 
In perturbation theory the perturbative vacuum
persists in any order and the quark condensate (as well as dynamical mass)
are identically zero.
It is therefore inconsistent
to invoke of constituent quarks along with perturbative
one gluon exchange. If
one invokes constituent quarks, then one necessarily
assumes the spontaneously broken mode of chiral symmetry, where
the Goldstone boson field is required by the Goldstone theorem
and the flavor-octet axial current conservation in the chiral limit
implies the coupling of Goldstone bosons and quasiparticles 
\cite{NJL,MG}\footnote{To avoid any confusion, one should not
mix the one gluon exchange interaction between constituent quarks
with its nonrelativistic 
spin-spin force, with the nonperturbative resummation
of the gluonic exchanges between current quarks by solving the
Dyson-Schwinger and Bethe-Salpeter equations, which could provide
chiral symmetry breaking (this is one of the possibilities which
is presently discussed \cite{RW}) and which will automatically lead to
GBE between the quasiparticles in baryons upon t-channel iterations
\cite{GV}. In this approach, however, the $U(1)_A$ problem persists and the
origin of confinement is unclear. In this case the underlying
mechanism for the $\pi - \rho$ splitting is the same like
in Nambu and Jona-Lasinio model \cite{Reviews}
and has nothing to do with the
nonrelativistic spin-spin force between quarks.}.
The fact that the typical momentum of valence
{\it current} quarks in the nucleon is small, $\sim 100 - 200$ MeV,
i.e. well below the chiral symmetry
breaking scale, $\Lambda_\chi \sim 1$ GeV, implies that the low-energy
characteristics of baryons, such as their masses, should be formed
by the nonperturbative QCD dynamics that is responsible for the
chiral symmetry breaking and confinement, but not by the perturbative
QCD degrees of freedom which become active at much higher momentum
scales. The effective meson exchange interaction between valence
quarks in baryons arises from the nonperturbative
t-channel iterations
of the QCD gluodynamics which triggers the breaking of chiral
symmetry and which is responsible for the low-lying meson
structure \cite{GV}. This is a simple consequence of crossing symmetry:
if one obtains the pion as a solution of the Bethe-Salpeter
equation in the quark-antiquark s-channel, then one inevitably
obtains pion exchange in the quark-quark systems as a result
of iterations in the t-channel. What is important is
that these t-channel iterations enormously reinforce the
bare (gluonic) vertex  in the GBE channel (which is due to the
 antiscreening) at small momenta. This antiscreening results in
the pole that occurs
at $q^2=0$. This pole "explosion"
explains the role of the GBE interaction at low momenta,
which dominates the low-energy baryon physics. 
Generation of the dynamical mass and the
exchange by Goldstone bosons between quasiparticles in baryons are 
synchronous phenomena based on chiral symmetry breaking
and cannot be separated from each other.
\\

There are many independent indications
from spectroscopy that show that the physics in the heavy
quark sector (where the chiral symmetry is absent and the 
confinement plus OGE picture is a relevant one)
is very different from the light quark sector. For instance,
the hyperfine (spin-spin) splittings in charmonium are
of the order of 3\% of the hadron mass, i.e. they indeed
represent a small perturbation. In contrast, the spin-spin
force in light baryons should be very strong as it provides
splitting at the level of 30\% of hadron mass ($N - \Delta$)
splitting. The color-magnetic spin-spin interaction in the
heavy quark systems has a clear origin as a small nonrelativistic
$v^2/c^2$ correction to the leading Coulomb force of OGE
interaction, and, as it is well known from the positronium physics
 (which is similar) provides a small
$\sim \alpha^4$ spin-spin splitting \cite{LANDAU} and in the present case
the values of $v^2/c^2$ , $\alpha^4$ and the experimental splitting
are all consistent to each other. In the light quark systems, the
light current quarks with their tiny mass are ultrarelativistic.
In this case the perturbative gluon - quark vertex to a good
approximation 
conserves helicity (to be contrasted to heavy quark - gluon vertex), which
implies that the spin dependence of OGE interaction vanishes 
in the present case \footnote{ I repeat, once one considers the standard
one gluon exchange perturbative force, then one assumes that
we are in the perturbative regime of QCD, i.e. on the top of
perturbative vacuum. Hence one can use only original (current) quarks
of QCD in the present case.}. This is in obvious conflict with
the large empirical hyperfine $N - \Delta$ splitting and implies
that the {\it perturbative} gluon exchange force cannot be its origin.\\

The lattice calculations 
indicate that the physics in the heavy quark sector 
 is very different from the light quark sector.
To these belong a recent analysis by Liu et al \cite{LIU}, showing
that the origin of the $N-\Delta$ splitting is not due to the
color-magnetic interaction, but inherently related with the dynamical
chiral symmetry breaking and meson-like exchange force. 
The most recent work of RIKEN BNL -
Columbia - KEK collaboration
\cite{SASAKI}, which for the first time accurately measured the
low-lying negative parity state\footnote{This can be considered
as a proof that N(1535) is a genuine three quark resonance,
but not a cusp due to the nearby $N\eta$ threshold and not
a quasibound state in the meson-baryon system.}
and also obtained a reliable signal
for the Roper  state,  indicates that the dynamics 
for baryons made of  heavy quarks (where the chiral symmetry and
GBE-like force are absent), in which case the spectrum
indeed looks like in Fig. 1, is very different from the real 
pattern in nature on Fig. 2, which is
close to the chiral limit.\\  

\section{"The Spin-Orbit Problem is not Solved"}

Here the argument made by Isgur is as follows. The empirical
spectra of L=1 light baryons and mesons show no
significant spin-orbit splittings. A scalar confining interaction
implies a spin-orbit force due to Thomas precession, which should
be cancelled by another spin-orbit force in both baryons 
and in mesons.
Such an additional spin-orbit force is supplied by a strong
one-gluon exchange interaction, while within the GBE model 
for baryons there is no source to counterbalance
the Thomas term.\\

This argument is based on the naive extrapolation of heavy quark
physics into the light quark sector. In the heavy quark systems,
like charmonium or bottomonium, the most important dynamics is
indeed due to the string-like confining force at large
distances and a small perturbative gluon exchange correction
at short ones. In this case a heavy quark practically constantly
"sits" on the end of the string because a quantum-mechanical
fluctuations of this quark into other one plus quark-antiquark
pair (meson) are suppressed by the factor $1/M_Q$ 
(see footnote 14)
and vanish
in the heavy quark limit. This suppression factor comes from the
meson propagator.\\

A relativistic rotation of the string implies the Thomas
precession, which is a pure kinematical effect related to
successive Lorentz transformations. This Thomas precession 
gives rise to 
a spin-orbit interaction. Note that for this effect to be
operative it is necessary to have {\it the same} particle on the
end of the string at the successive moments $t_1, t_2$ and $t_3$.
For the heavy quark this condition is 
indeed approximately fullfilled.\\

\begin{figure}
\psfig{file=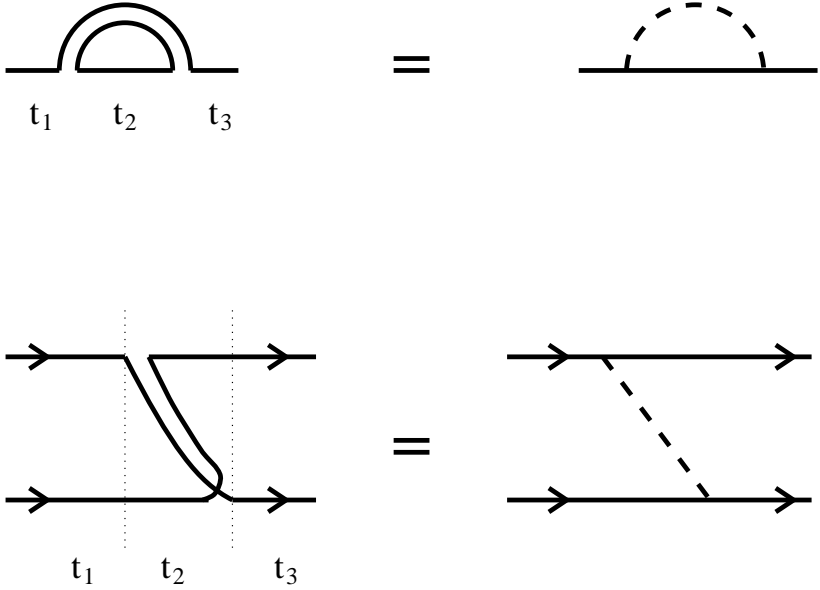}
\caption{Pion fluctuation of the light constituent quark.}
\end{figure}

In the light quark systems this condition is not satisfied, however. 
This is because 
quantum mechanical fluctuations of the light valence
quark into the {\it other} quark and the light meson are not
suppressed and become the most important effect. Within the
quantum field theory such a fluctuation corresponds to
materialization from the vacuum of the quark, which becomes the
valence  instead of the initial one, see Fig. 3. Because
in the present case there is no big gap between the negative
energy levels of the Dirac sea and the positive energy of the
valence quarks, this process is intensive. This implies that
at the successive moments $t_1, t_2$ and $t_3$ one has
predominantly {\it different} quarks on the end of the string, 
though with exactly the same
color. If quarks are
different, the Thomas precession {\it cannot be applied}.
In addition the spin of the quark at the moment $t_2$ is predominantly
polarized just in opposite direction compared to the moments
$t_1$ and $t_3$ as the pion-quark vertex is of spin-flip nature \cite{HRG}.
Thus, at $t_2$ the spin-orbit Thomas term is
of opposite sign compared to that at $t_1$ and $t_3$\footnote{
One may speculate whether the loop fluctuation also
affects the spin-spin force from the GBE between different quarks.
The pion-quark vertex is of spin- and isospin-flip nature
which means that the loop contributions to the one pion exchange,
where exchanged pion is attached to quark within a loop,
produces the same
 operator
$-{\vec\tau}_i\cdot {\vec\tau}_j {\vec \sigma}_i \cdot {\vec \sigma}_j$
as one pion exchange without loop.}. This
qualitative discussion suggests that the Thomas spin-orbit
force should be strongly suppressed in the light quark 
systems, both in mesons and 
baryons\footnote{A detailed formal extension of this qualitative
discussion will be published elsewhere.}.\\

If so, the spin-orbit force from the OGE which should be
very strong as it is fixed by large $N-\Delta$ and $\pi -\rho$
splittings within the naive OGE model 
(combined with constituent quarks)
completely destroys both
baryon and meson spectra as it supplies splittings of
{\it hundreds} MeV.\\

Based on the view of the near perfect cancellation
of the very large, but opposite in sign, LS forces from
Thomas precession and OGE interaction in P-wave light mesons
(which, according to Isgur, should be of 
the same origin like in light P-wave baryons)
, Isgur interpolates their matrix elements
between  the light and the heavy quarkonia to the heavy-light mesons
and  predicts a dramatic and large inversion of the
spin-orbit splittings in the heavy-light P-wave mesons, where the
data were absent (for details see ref. \cite{Isg}).
This prediction has recently been checked by two independent lattice
groups \cite{Wol} and has been ruled out. Not only does
this prediction deviate from the data by a few hundreds MeV, but
its sign is opposite!\\

In fact there do appear spin-orbit forces in the GBE model from
the second iteration of the interaction \cite{GR,G,RB}, which
correspond to spin-orbit force from vector- and scalar-meson
exchanges. Different meson exchanges provide the spin-orbit force
with opposite signs in baryons \cite{G,RB}, which suggests that
the net spin-orbit force should not be large, which is compatible
with the small 10-50 MeV LS-splittings observed in L=1 light and
strange baryons.\\

To conclude this section we stress that it is incorrect
to identify the linear confining interaction between two
heavy static sources, that is indeed established, with an
effective confining interaction between the {\it quasiparticles}
in the light quark systems.\\

\section{"Baryon Internal Wave Functions are Wrong"}

Here Isgur's argument is that while the OGE model yields a
mixing of the spin $S=1/2$ and $S=3/2$ states in
the $N(1535)$ and $N(1650)$ baryon wave functions, which is compatible
with the big observed $N(1535) \rightarrow N\eta$ branching
ratio and the small  $N(1650) \rightarrow N\eta$
one, the GBE model should fail to do so.\\

This mixing above is provided by the tensor force component
of the quark-quark force and crucially depends on its sign,
while the masses of baryons are not strongly sensitive to
this tensor force. Within the GBE picture there are two
sources for tensor force: pion-like exchange and rho-like exchange
mechanisms.
Both of these exchanges supply a spin-spin force {\it with the
same sign}, while their tensor force components have 
{\it opposite} signs \cite{G}. This implies that the net tensor
force should be rather weak compared to the strong spin-spin
force, in agreement with phenomenology. In ref. \cite{GR} only
a  $\pi$-exchange tensor force was used for an estimate.\footnote{
It has nevertheless been stressed there: "Any vector-octet-like exchange
interaction component between the constituent quarks, would also
reduce the net tensor interaction at short range as the contributions
to the tensor interaction from pseudoscalar and vector
exchange mechanisms tend to cancel, whereas they add in the case
of the spin-spin component. These modification of the tensor interaction
at short range may even lead to a sign change of the matrix element."} 
 Its
strength has not been correlated with the strength of the
spin-spin force, which is fixed by the hyperfine splittings.
As soon as the corresponding $\rho$-exchange like tensor force
is added, the mixing becomes qualitatively different. 
The flavor dependent tensor force component of the
two-pion exchange interaction (which is $\rho$-like)
is, in the range relevant
for the baryon wave functions stronger than that of the
one-pion exchange interaction \cite{RB}, and therefore the net
tensor force, while weak, does have the sign opposite
to that of pion exchange. The sign is then that, which
is {\it favored} by the empirical mixing of the
negative parity multiplets. Note that in the modern fits of baryon
spectra both $\pi$-like and $\rho$-like exchanges are taken into
account \cite{PANIC,W,G}.\\

There are several indications that the $\rho$-like
tensor force should dominate over the $\pi$-like in P-wave
baryons. The analysis of the L=1 spectra and of the mixing
angles for the flavor-dependent interaction \cite{Collins}
reveals that the tensor force that mixes $S=1/2$ and $S=3/2$
components should have a sign of the $\rho$-exchange tensor
interaction\footnote{While this fact is not explicitly discussed
in that paper, it follows from the mixing angles presented
in Table 3 therein.}. 
With the flavor-dependent spin-spin and tensor force, with
the matrix element being adjusted to provide the best $\chi^2$
fit to baryon masses, a parameter-free prediction
for mixing angle was obtained, which ideally fits
the observed $\pi$ and $\eta$
decays branches for  $J=1/2$   and $J=3/2$ $L=1$ $N^*$ baryons,
discussed in  Isgur's paper \cite{Isgur}.
That work definitely shows that the fit of the
observed $L=1$ spectra prefers a flavor-dependent interaction
between quarks.\\ 

This is perfectly consistent with the recent systematic $1/N_c$
analysis of both masses and mixing angles of L=1 nonstrange
baryons \cite{Carone}. The result of this paper may be 
summarized as follows: both masses and mixing angles extracted
from the strong and electromagnetic decays are compatible with
the idea that the effective quark-quark interaction is of meson
exchange 
form, while they are not compatible with the flavor independent
gluon exchange hyperfine interaction. In particular the data
{\it require} the significant contribution of the operator that
contains a flavor-dependent tensor force ($O_3$), while the
contribution of the operator which represents a flavor-independent
tensor force ($O_8$) is compatible with 0. Note that in the
present analysis the contribution of different operators is 
systematically weighted
with the $N_c$-dependent factor, which is absent in other more
phenomenological analyses. The study of the $\pi N$ phase shifts
\cite{Sato} also reveals that the spin-spin force between quarks
should be of pion-exchange type, while the tensor force component
should be of just opposite sign\footnote{Those authors actually 
conclude that the spin-spin force should be of pion-exchange type
while the tensor force should of gluon-exchange type, which
would be rather strange. But it is
easy to see from their expressions that the same result will be
obtained if one changes the sign of the single $\pi$-exchange
tensor force to the opposite one.}.\\

This should not be construed as a claim that a simple $QQQ$ 
main component
of the baryon wave function alone will be able to explain the
variety of strong and electromagnetic decay data. The baryon
wave function contains in addition further Fock components,
$QQQ+meson,...$. The coupling of these higher Fock components
will be very important for strong decays in the case when the
energy of the resonance is close to the corresponding threshold.
In this case the energy denominator, which determines
a role of the higher Fock component in the given reaction,
e.g. in $\gamma N ~or~ \pi N \rightarrow N(1535) \rightarrow N \eta$,
becomes very small and  the otherwise insignificant
$QQQ\eta$ component of the $N(1535)$ wave function becomes
important. This should be a significant reason for why the
$\eta$-decay branch is anomalously big in the case of $N(1535)$.
Note that within the chiral constituent quark model this mechanism
is very natural, while there are no meson components in the baryon
wave function within the OGE model. 
\\

Similar arguments can be applied to explain an anomalously
large $\Lambda(1405) - \Lambda(1520)$ spin-orbit splitting,
because the $\Lambda(1405)$ is below the $\bar KN$ threshold and can
be viewed as $\bar K N$ bound state \cite{Dalitz}. If correct, it
would simply mean that both coupled $QQQ$ and $QQQK$ components
are significant in the present case and there is no contradiction with
the flavor singlet $QQQ$ nature of these baryons, which  in any
case are LS partners with respect to their main $QQQ$ component.
The alternative explanation
of the latter extraordinary large LS splitting would be that there
is some rather large spin-orbit force specific to the flavor
singlet state only \cite{GR,Collins}, which is also not rulled
out, while it is clear that OGE cannot supply such a flavor
dependent LS force. The mixing pattern of singlet and octet
components that is obtained with the flavor-dependent interactions
in ref. \cite{Collins} better describes the strong decays
of $\Lambda(1405)$ than that one obtained with the 
flavor-independent interaction.
\\

To conclude this section one should stress that 
in the present state of the art it is
premature to judge on the effective $QQ$ interactions
from the  strong decays. This is, in particular,
 because the excited states are treated within the
quark model as bound states, rather than as resonances, and
an incorporation of the $QQQ\pi$,... continuum components
coupled to the principal one $QQQ$ is vital for strong decays.
A real test of any constituent quark model
 beyond spectroscopy (i.e. also of their wave
functions) can reliably nowadays
be performed only for the ground state observables. 
Such a task has just   been completed
for the chiral constituent quark model \cite{Wag}. 
Starting out from the wave functions obtained in ref. \cite{GPVW},
which represent the eigenstates of the mass operator
 of the
manifestly covariant point form of relativistic quantum
mechanics, one has calculated nucleon e.m. formfactors
performing relativistic boost transformations 
\cite{Klink}. The  {\it parameter free} predictions
for proton and neutron electric and
magnetic formfactors as well as charge radii and magnetic
moments turned out very satisfactory and practically explain
existing data
 (e.g. within the experimental
error bars for proton and neutron charge formfactors).
It is also demonstrated  that using the same wave functions
but a nonrelativistic framework for  calculating these
observables, the  formfactors and charge radii
result completely differently, deviating
 by 1-2 orders of magnitude. From this comparison one
can conclude, in particular, that the proper inclusion
of Lorentz boosts is crucially important.
In view of all that
nonrelativistic calculations within the
constituent quark models (or  similarly within bag models), 
appear very questionable. 
This is especially  true with regard to  strong decays.

\section{"Mesons are Disaster"}

There are several  arguments suggested by Isgur in this section.
The first one is that while the GBE (or, generally, meson
exchange like interactions) may be possible in baryons, such are 
impossible between valence quark and antiquark in mesons
(e.g. in $u\bar d$ pair) within the quenched approximation to
QCD, thus suggesting that meson and baryon spin-dependent 
interactions must
have totally different physical origins which is very difficult
to arrange.\\

This question has been addressed in detail recently \cite{GV}.
I will briefly summarize here the main conclusions. One needs
a nonperturbative gluonic interaction between quarks in QCD
to provide chiral symmetry breaking. A good candidate is 
the instanton-induced 't Hooft interaction \cite{Hooft,DP,Shuryak}.
When this nonperturbative gluonic interaction breaks chiral
symmetry, i.e. generates at low momenta the constituent mass $m$ of quarks, 
it also automatically supplies a strong attractive
interaction in the pseudoscalar-isovector quark-antiquark system -
pions - which makes them anomalously light, with zero mass in
the chiral limit. This is how the pions appear as the 
Nambu-Goldstone bosons of the spontaneously
broken chiral symmetry. This mechanism is well illustrated 
by the Nambu and Jona-Lasinio model \cite{NJL}. While there
is a strong attractive interaction in the pseudoscalar-isovector
quark-antiquark system, the interaction is {\it absent} to leading
order in vector mesons, which means that masses of vector mesons
should be approximately $2m$, which is well satisfied empirically,
$\mu_\rho \simeq \mu_\omega \simeq 2m$. The implication is that the
$\pi - \rho$ mass splitting is not due to the perturbative 
color-magnetic interaction between spins of constituent quarks in
$\pi$ and $\rho$, but entirely due to the fact that the QCD
Lagrangian posseses a chiral symmetry which is dynamically
broken in the QCD vacuum. Note that the 't Hooft interaction
also naturally solves the $U(1)_A$ problem, explaining thus
why $\eta'$ is heavy, contrary to $\pi$. This problem cannot
be solved by the OGE interaction as a matter of
principle. 
\\

\begin{figure}
\psfig{file=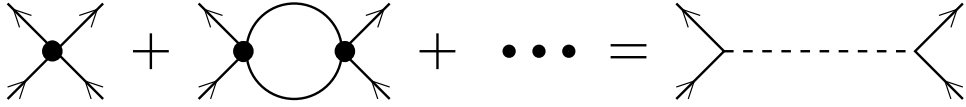}
\caption{Iteration of the instanton-induced 't Hooft interaction
(or some other gluonic interaction which is responsible for the
chiral symmetry breaking in QCD)
in the qq t-channel in baryons. Black filled circle means a bare gluonic vertex.}
\end{figure}

The Nambu and Jona-Lasinio mechanism of chiral symmetry breaking
(and hence of $\pi - \rho$ splitting) is the most general one. It
only exploits the fact that the quark-gluon interaction in QCD respects
chiral symmetry. In fact one does not need to assume  that it is the 
instanton-induced interaction which provides chiral symmetry
breaking.\\

When that nonperturbative gluonic interaction between quarks,
which is responsible for chiral symmetry breaking in QCD,
is iterated in the qq t-channel in baryons, it inevitably leads
to poles which correspond to a GBE interaction in quark-quark
pairs, see
Fig. 4. This is a typical antiscreening behavior, the
interaction of two quarks in baryons is represented by a bare
gluonic vertex at large momenta transfer (i.e. at very small
distances), but it blows up at small momenta in the channel
with GBE quantum numbers, explaining thus a distinguished role
of the latter interaction in the low-energy regime.
Thus
the GBE interaction in baryons is in fact an effective representation
of the t-channel ladders, which strongly reinforce a bare gluonic vertex
at low-momentum
transfer in the GBE channel. 
Since the typical momentum of valence current quarks in
baryons is well below the chiral symmetry breaking scale,
these interactions dominate (see Introduction).
This suggests that the origin
of the hyperfine splittings in both the low-lying mesons and baryons
is intrinsically the same - it is the nonperturbative gluonic interaction
between quarks which is responsible for chiral symmetry breaking
in QCD - which, however, reveals itself 
differently in mesons and baryons.\\

In Fig. 2 of his paper Isgur shows an evolution of the hyperfine 
splittings in mesons starting from the heavy quarkonium to $\pi - \rho$
mass splitting, arguing that it supports "a smooth evolution
of the wave function ... convoluted with the predicted $1/m_Q^2$
strength of the OGE hyperfine interaction". This 
figure is misleading.
Even if one takes a naive view that the $\pi - \rho$ splitting
 is due to OGE spin-spin force between the constituent quarks, 
one 
cannot explain why
the pion is very light, but $\eta, \eta'$ are heavy 
since this  spin-spin force  {\it must provide} 
the same strong
attraction also in $\eta,\eta'$, or, 
in other
words, one cannot explain in this approach
why $\pi - \rho$ mass splitting is big but 
$\eta' - \frac{\sqrt{2}\omega + \phi}{\sqrt{3}}$
mass splitting is even opposite in sign.
This fact alone rules out this naive mechanism of the
light pseudoscalar-vector meson splittings.
\\

The Fig. 3a of Isgur's paper claims to support the same idea
using the heavy-light mesons. According to Isgur this figure
illustrates the $1/m_q$ behaviour of the OGE spin-spin force
splittings, where $m_q$ is light quark mass. Again, this figure is
as misleading
as the previous one, as it does not show really
$\it clean$\footnote{The transition from the $B-B^*$ and $D - D^*$
systems to $K - K^*$ and $\pi - \rho$ is dubious one as
in the former case the system is indeed heavy-light, while
in the latter it is light-light.} examples which rule out this 
behavior. To these
belong the hyperfine $D - D^*$ splitting of 141.4 MeV and
$D_S - D_S^*$ one of 143.8 MeV. In the former case one has
$\bar c - u$ or $\bar c - d$ system, while in the latter one the
$u$ or $d$ quark is substituted by a strange one. One obviously
observes an absence of the $1/m_q$ behaviour as the constituent
light and strange masses differ by 30 - 50 \%. Exactly the same
situation takes place in B-meson, compare the hyperfine
 $B-B^*$ splitting of 45.7 MeV with that one of $B_S - B_S^*$,
of 47.0 MeV. One can then conclude that while there is indeed
the $1/m_Q$ scaling with respect to heavy quark mass $m_Q$ in
the heavy-light systems, which follows from the heavy quark
limit (symmetry) in QCD \cite{IW}, there is {\it no} similar
behavior with respect to light quark component of the heavy-light
systems. Similar objections can be raised against Fig. 3,b of 
Isgur's paper.\\

Needless to say that the Isgur's statement about the large
$\pi - \rho$ mass splitting as originating in the nonrelativistic
spin-spin force component of OGE {\it perturbation} comes
into conflict with the current algebra and all subsequent
developments in QCD which show unambiguously that the low
mass of pion, which is approximate Goldstone boson 
(and which is of course a quark-antiquark system), 
is due to the chiral symmetry dynamical breaking
in the QCD vacuum. Even if one assumes that the dynamical
chiral symmetry breaking comes from the nonperturbative
resummation of gluonic exchanges by solving the Dyson-Schwinger
equation for the quark Green function and the low mass 
of pion from a simultaneous solution of
the Bethe-Salpeter equation for the quark-antiquark system
with the gluon-exchange kernel,  the low mass of the pion
(and $\pi - \rho$ splitting) in this
case has nothing to do with the quark-antiquark 
nonrelativistic spin-spin force (see footnote 5).
\\

However, it is indeed the case that the small
hyperfine splittings in the heavy quarkonia are due to the
nonrelativistic color-magnetic spin-spin force stemming
from the small OGE perturbation. This mechanism, while
important at the botom and charm quark mass scales, dies
out in the region  between the charm and strange quark
scales (it vanishes in the chiral limit). On the
other hand near the chiral limit the splittings are due to
the chiral symmetry dynamical breaking, which, in turn,
should decrease with increasing the current quark mass. 
It then follows that the smooth evolution of the 
splittings shown in Figs. 2 and 3 of Isgur's paper, is
due to a superpositions of these two pictures.\\

Next Isgur argues that the annihilation graphs, see his Fig. 4,
which are possible only in the isoscalar channels in mesons,
but not possible in the isovector ones and which violate the
OZI rule should produce strong splitting in $\rho - \omega$
system  as well as a strong mixing of the $\bar u u, \bar d d$ and 
$\bar s s$ components in $\omega$ and $\phi$,
if one assumes that the GBE graphs between quarks in
baryons induce a $\Delta - N$ splitting. 
This problem has a very simple resolution
if one assumes that the 
instanton - induced 't Hooft interaction is the most important one.
These annihilation graphs do contribute
in the pseudoscalar mesons and provide the solution of the 
$\pi - \eta - \eta'$ puzzle. 
However, there are {\it no} such graphs from 't Hooft interaction
in vector mesons \cite{Hooft2,Shuryak}. It is this pecuilarity which explains
the completely different mixing of singlet and octet components
in the pseudoscalar and vector mesons, which is unnatural
in the former case and natural in the latter one \cite{Hooft2}.
As explained in the beginning of this section
the GBE interaction in the quark - quark systems 
(to be contrasted to the quark - antiquark ones)
can be regarded a result of the t-channel iterations of 
the same (like in mesons) bare 't Hooft vertex.\\

\section{ "The Connection to Heavy Quark Baryons is Lost"}

Here Isgur again uses his  Figs. 2 and 3 for argumentation.\\

While it is correct that around the heavy quark limit the OGE
mechanism is indeed important for small hyperfine splittings,
the light quark limit (chiral limit) is just opposite one
in QCD and implies completely different dynamics, inherent
in QCD. There are no doubts that it is a chiral dynamics,
i.e. dynamics of massless quarks in external gluonic
fields
which becomes the most important phenomenon in this case.
As argued in the previous section no  conclusions
can be obtained from these figures, which ignore 
well known empirical data. \\

What then the dynamics is, that is responsible for the heavy - light
systems, and, in particular, heavy-light baryons\footnote{It is,
unfortunately,
an incorrect statement in \cite{Isgur} that the exchange
by heavy-light meson (e.g. $D,D^*$) between heavy-light quark pairs
in baryons \cite{GLRI} should produce $1/M_Q^2$ scaling, in 
contradiction with the heavy quark limit. Naively,
from
the covariant  meson propagator one would indeed
obtain scaling $1/M_Q^2$. This scaling comes from both
the positive energy solution propagating forward in time
and the negative energy solution propagating backward in time.
However, the "heavy meson exchange"  viewed as in
Fig. 3  with the Z-like part
made of the light quark line only and with the heavy quark propagating
only forward in time scales as $1/M_Q$.
The heavy quark Z-like line , which would correspond
to the propagation of the heavy quark 
 backward in time, is suppressed by heavy quark
mass.}
is an open question (it cannot be excluded that
in the  case of baryons both meson-like dynamics and perturbative
QCD corrections are equally important).
At least, what is known, the prediction of the spin-orbit
splittings in heavy-light mesons \cite{Isg}, based on the scalings of
Figs. 2 and 3 and OGE, turned out in dramatic disagreement with
the very recent lattice results \cite{Wol}. 
\\

Returning then to the question of the splitting
of the $\Lambda(1405) - \Lambda(1520)$ multiplet
and its charm analog $\Lambda_c(2954) - \Lambda_c(2627)$ there
is no
objection to their dynamical similarity, which suggests
that the 
$\Lambda(1405)$ should have a large  $QQQ$ component.
As explained in the  section 3, it does not
contradict the idea that there is an appreciable higher Fock
component $QQQ\bar K$, which provides an anomalously large
$\Lambda(1405) - \Lambda(1520)$ splitting. The other possibility,
that there exists some spin-orbit force, which is specific to the
flavor - singlet state only is also not ruled out, while
it is clear that OGE, which is flavor independent, cannot
provide such a spin-orbit force.\\

\section {Conclusions}

In "Conclusions" Isgur raises a few conceptual objections.
The first one is about a double-counting problem since a
theory which uses both constituent quarks and Goldstone bosons
has "both fundamental Goldstone bosons and quark-antiquark
bound state Goldstone bosons".\\

This objection is obviously based on misunderstanding of the
low-energy effective theory. There is no fundamental Goldstone
boson field in QCD. The pion as a Goldstone boson is of course
a system of quarks and antiquarks and has entirely dynamical
origin \cite{NJL}. It arises naturally as a deeply bound state
from the corresponding microscopical quark-gluon nonperturbative
interaction in QCD, e.g. the instanton - induced one. When one
applies the same Lagrangian (which does not contain any pion
field!) in baryons and iterates it in the qq t-channel,
one arrives at the pole contribution which corresponds to
GBE between quarks in baryons \cite{GV}. This is a simple
consequence of crossing symmetry: if one obtains pion as
a solution of the Bethe-Salpeter equation in the quark - antiquark
s - channel, then one inevitably obtains a pion - exchange in
the quark - quark systems as a result of iterations in the qq t-channel.
There is no fundamental pion-exchange between quarks as there is
no fundamental pion field in QCD. The pion exchange is not more than
an effective representation of the t-channel ladders in the low-energy
and low-momentum regime where these ladders become important.\\

The second problem "is that it is not legitimate to treat the
quark-Goldstone boson vertex as pointlike". In fact that was never
suggested  and instead it has been insisted, in all papers, that
the finite size of both constituent quarks and pions provides
a smearing of the otherwise contact short-range spin-spin 
quark-quark force. It is this smeared short-range part of GBE
interaction that is crucially important for splittings in baryons.
Indeed, the results crucially depend on the smearing parameter
\cite{GPVW}, that should be originated from the intrinsic
structure of pion and also from unknown nonlinear behavior of
the effective chiral Lagrangian \cite{Glozman}.
\\

The third objection that "there is no obvious rationale
for truncating the tower of meson exchanges ..." was
addressed in the section 3. Obviously all mesons
should contribute. An important issue, however, is that the
spin-spin force from $\pi$, $\rho$ or $a_1$ meson exchanges in
quark-quark system has exactly the same flavor-spin structure
and sign at short range, which is crucial for baryon spectroscopy,
so they only enhance the effect of each other, while the tensor
and spin-orbit forces from different meson exchanges interfere
destructively in baryons \cite{G}, which explains a significant
spin-spin force and at the same time rather weak net tensor and
spin-orbit forces, which is suggested by empirical baryon spectra.
Nevertheless, the importance of different meson exchanges is
different and is determined by the position of the corresponding
pole at the unphysical time-like momenta in the quark-quark system
(i.e. in baryon). The closer
a pole is to the space-like region, which determines the quark-quark
interaction, the more important the given meson exchange is.
 The pion pole is located just at
the origin of the space-like axis and thus strongly influences the
quark-quark interaction in baryons in the regime where momentum transfer
is not large.\\

In summary the idea of the GBE
model in baryons is not that there is no perturbative gluon exchange
in QCD and, in particular in light baryons and mesons, but that
such contributions cannot be significant  for the
{\it low-energy} observables such as masses, where
the dynamics is driven by  nonperturbative phenomena among which
the crucially important are dynamical chiral symmetry breaking
and confinement. The  importance of the GBE flavor-dependent
spin-spin force is not only conceptually substantiated,
but it is also strongly supported by the fact that once one
extracts the pion-quark coupling constant from the well known
pion-nucleon one, regularizes the $\pi q$ vertex with the
cutoff of the order $\Lambda_\chi \sim 1$ GeV and solves the
(semi)relativistic 3-body equations exactly, 
the $N-\Delta$ splitting turnes out
of the order 300 MeV (or larger!). At the same time the
Roper state is shifted down below the negative parity multiplet.
The addition of any sizable
phenomenological
 color-magnetic force between the
constituent quarks explodes the baryon spectra \cite{GPPVW}. 
\\

\section{Acknowledgement}

I am grateful to D.O. Riska for a careful reading of the manuscript.


\begin{thebibliography}{99}
    \bibitem{Isgur} N. Isgur, Phys. Rev. {\bf D62} 054026 (2000);
     nucl-th/9908028 (versions 1 and 2).
    \bibitem{GR} L. Ya. Glozman and D. O. Riska, 
     Phys. Rep. {\bf 268}, 263 (1996).
    \bibitem{GPVW} L. Ya. Glozman, W. Plessas, K. Varga, and R. Wagenbrunn,
     Phys. Rev. {\bf D58} 094030 (1998)
    \bibitem{PANIC} L. Ya. Glozman, Plenary talk given at PANIC 99, 
     Nucl. Phys. {\bf A663,664} 103c (2000).
    \bibitem{CKP} J. Carlson, J. Kogut, and V.R. Pandharipande,
     Phys. Rev. {\bf D27}, 233 (1983)
    \bibitem{SB} B. Silvestre-Brac and C. Gignoux, Phys. Rev. {\bf D32}, 
    743 (1985)
    \bibitem{KN} G. Karl and V.A. Novikov,
     Phys. Rev. {\bf D51}, 5069 (1995)
    \bibitem{CI} S. Capstick and N. Isgur,
     Phys. Rev. {\bf D34}, 2809 (1986)
    \bibitem{IK}  N. Isgur and G. Karl,
     Phys. Rev. {\bf D18}, 4187 (1978);{\bf D19}, 2653 (1979)
    \bibitem{Capstick} S. Capstick and W. Roberts, nucl-th/0008028
    \bibitem{Aoki} S. Aoki et al, Phys. Rev. Lett. {\bf 82}, 4592 (1999)
    \bibitem{Sku} J. I. Skullerud and A. G. Williams, hep-lat/0007028
    \bibitem{NJL} Y. Nambu and G. Jona-Lasinio, Phys. Rev. {\bf 122}
    345 (1961); {\bf 124} 246 (1961)
    \bibitem{RW} C. D. Roberts and A. G. Williams, Progr. Part. Nucl. Phys.,
    {\bf 33}, 4777 (1994) 
    \bibitem{Reviews} U. Vogl and W. Weise, Progr. Part. Nucl. Phys.
    {\bf 27} 195 (1991); S. P. Klevansky, Rev. Mod. Phys.,
    {\bf 64} 649 (1992); T. Hatsuda, T. Kunihiro, Phys. Rep. {\bf 247}
     221 (1994).
    \bibitem{GV} L. Ya. Glozman and K. Varga, Phys. Rev. {\bf D61},
     074008 (2000) 
    \bibitem{LANDAU} L. D. Landau and E. M. Lifshitz, Course of
    Theoretical Physics, vol. 4, Pergamon Press, 1982 
    \bibitem{LIU} K. F. Liu et al, Phys. Rev. {\bf D59} 112001 (1999)
    \bibitem{SASAKI} S. Sasaki, hep-ph/0004252
    \bibitem{MG} A. Manohar and H. Georgi, Nucl. Phys. {\bf B234}
    189 (1984). 
    \bibitem{DP} D. Diakonov and V. Yu. Petrov, Nucl. Phys. {\bf B272}
    457 (1986). 
    \bibitem{HRG} L. Hannelius, D.O. Riska and L. Ya. Glozman, 
     Nucl. Phys. {\bf A 665} 353 (2000)
    \bibitem{Isg} N. Isgur, Phys. Rev. {\bf D57} 4041 (1998)
    \bibitem{Wol} R. Lewis and R. M. Woloshin, hep-lat/0003011
    \bibitem{W} R. Wagenbrunn, L. Ya. Glozman, W. Plessas, K. Varga,
    Nucl. Phys. {\bf A 663,664} 703c (2000)
    \bibitem{Collins} H. Collins and H. Georgi, Phys. Rev. {\bf D59}
     094010 (1999)
    \bibitem{Carone} C. E. Carlson et al, Phys. Rev. {\bf D59} 114008 (1999);
     C. D. Carone, hep-ph/0004216
    \bibitem{Sato} T. Yoshimoto, T. Sato, M. Arima, T.-S.H. Lee,
     nucl-th/9908048
    \bibitem{G} L. Ya. Glozman, Surv. High Energy Phys., {\bf 14} 109 (1999);
     hep-ph/9805345
    \bibitem{RB} D. O. Riska and G. E. Brown, Nucl.Phys. {\bf A653}
     251 (1999) 
    \bibitem{Dalitz} R. H. Dalitz and A. Deloff, J. Phys. {\bf G17}
     289 (1991) and references therein.
    \bibitem{Wag} R. F. Wagenbrunn et al, to be published
    \bibitem{Klink} W. H. Klink, Phys. Rev. {\bf C58} 3587 (1998);
     {\bf C58} 3617  (1998)       
    \bibitem{Hooft} G. 't Hooft, Phys. Rev. {\bf D14} 3432 (1976)
    \bibitem{Shuryak} T. Sch\"afer, E. V. Shuryak, Rev. Mod. Phys.,
     {\bf 70} 323 (1998)
    \bibitem{Hooft2} G. 't Hooft, hep-th/9903189
    \bibitem{GLRI} L. Ya. Glozman and D. O. Riska, Nucl.
       Phys. {\bf A603} 326 (1996) [E: ibid {\bf A620} 510 (1997) 
    \bibitem{IW} N. Isgur and M. B. Wise, Phys. Lett. {\bf B232} 113 (1989);
     {\bf B237} 527 (1990)
    \bibitem{Glozman} L. Ya. Glozman, Phys. Lett. {\bf B459} 589 (1999)
    \bibitem{GPPVW} L. Ya. Glozman, Z. Papp, W. Plessas, K. Varga,
     R. Wagenbrunn, Phys. Rev. {\bf C57} 3406 (1998).
    \end{thebibliography}
    \end{document}